\documentstyle[preprint,aps,floats,epsfig]{revtex}
\tightenlines

\begin{document}

\renewcommand{\thefootnote}{\fnsymbol{footnote}}

\title
{Has the general two--Higgs--doublet model\\
unnatural FCNC suppression? An RGE analysis.
\footnote{Presented by G. Cveti\v c
at XXIth School of Theoretical Physics
Ustro\'n 1997, Poland; to appear in Acta Phys.~Pol.~B.}}

\begin{flushright}
BI-TP 97/46, YUMS 97-028, SNUTP 97-149\\
hep-ph/9711409
\end{flushright}

\vspace{0.5cm}

\centerline{{\large \bf
Has the general two--Higgs--doublet model}}
\centerline{{\large \bf unnatural FCNC suppression? 
An RGE analysis.\footnote{
Presented by G. Cveti\v c
at XXIth School of Theoretical Physics,
Ustro\'n, Poland, September 1997. To appear in Acta Phys.~Pol.~B.}}}

\vspace{1.cm}

\centerline{ G.~Cveti\v c}
\centerline{{\it  
Department of Physics, Universit\"at Bielefeld,
33501 Bielefeld, Germany}}

\vspace{0.5cm}

\centerline{S.S.~Hwang and C.S.~Kim}
\centerline{{\it Department of Physics, Yonsei University, 
Seoul 120-749, Korea}}
 

\renewcommand{\thefootnote}{\arabic{footnote}}

\begin{abstract}
There is a widespread belief that
the general two--Higgs--doublet model (G2HDM)
behaves unnaturally with respect to evolution
of the flavor--changing neutral Yukawa coupling
parameters (FCNYCP's) -- i.e., that the latter, although 
being suppressed at low energies of probes, in general
increase by a large factor as the energy of probes
increases. We investigate this, by evolving
Yukawa parameters by one--loop renormalization
group equations and neglecting contributions of
the first quark generation. For patterns of
FCNYCP suppression at low energies suggested by
existing quark mass hierarchies, FCNYCP's remain 
remarkably stable (suppressed) up to energies very 
close to the Landau pole. This indicates that G2HDM
preserves FCNYCP suppression, for reasonably
chosen patterns of that suppression at low energies.\\
PACS number(s):11.10.Hi; 12.15.Ff; 12.15.Mm; 12.38.Bx; 12.60.Fr
\end{abstract}
\setcounter{equation}{0}
\setcounter{footnote}{0}

\newpage
  
T.D.~Lee \cite{Lee} proposed already in 1973
the general model with two Higgs doublets [G2HDM,
referred often also as 2HDM(III)]. This is the model
with the most general Yukawa couplings of the two Higgs
doublets (to quarks). He investigated the model with
emphasis on CP--violating phenomena, and was followed
by Sikivie {\em et al.} \cite{Sikivieetal}.

Glashow and Weinberg \cite{GlashowWeinberg}
pointed out in 1977 that only those models
with two Higgs doublets which
possess specific discrete [or U(1)--type] family 
symmetries in the Yukawa sector [2HDM(II) and 2HDM(I)] 
have the property of complete
suppression of all flavor--changing neutral Yukawa coupling
parameters (FCNYCP's), 
and this is then true at any evolution energy.\footnote{
Low energy experiments show that those flavor-changing neutral 
coupling parameters (FCNCP's) which don't involve $t$ quark
are suppressed in nature at low energies $E\!\sim\!m_q$. For 
FCNCP's involving $t$, experimental 
evidence is not yet available.}
Since then, a large part of physics community 
has apparently regarded such models as being the only ones 
with a natural suppression of FCNYCP's. 
It appears that G2HDM's were then
not investigated until late eighties. 

During the last twelve years, 
there has been a moderate resurgence of works on G2HDM's
\cite{Brancoetal}--\cite{Savageetal}.\footnote{
Refs.~\cite{Brancoetal} and Refs.~\cite{Savageetal} 
concentrate largely on CP-violating and FCNC-violating
phenomena, respectively.}
These works investigate
{\em low} energy phenomena of G2HDM's, allowing nonzero but 
reasonably suppressed FCNYCP's at low (electroweak) energies.
Cheng, Sher and Yuan (CSY) \cite{CSY} and Antaramian, Hall
and Ra\v sin (AHR) \cite{AHR} proposed specific ans\"atze
for these parameters at low (electroweak) energies, 
motivated largely by 
the existing mass hierarchies of quarks. They pointed out 
that neutral scalars can then still 
be reasonably light ($M\!\sim\!10^2$ GeV) without violating 
available data on flavor--changing neutral processes. Thus, 
these two groups showed that FCNC suppression in G2HDM's 
is not intrinsically unnatural (i.e., ad hoc), since
it can be motivated by existing quark mass hierarchies.
Certainly, these arguments regard only the structure of the 
theory at low (electroweak) energies of probes. We note that
most of the phenomenological investigations in G2HDM's by 
other authors used essentially ans\"atze of these 
two groups.

We are not aware of any work on the
second aspect of G2HDM's, i.e., the question whether
these models really behave unnaturally with respect
to {\em evolution\/} of FCNYCP's. The present work attempts
to address this question, by investigating one--loop 
evolution of these parameters, for the cases when the
patterns of their values at low energies are described
by CSY--type of ans\"atze and modifications thereof. 

Yukawa interactions of quarks 
in G2HDM's in any $\rm{SU(2)_L}$-basis are
\begin{eqnarray}
{\cal L}^{(E)}_{\rm{G2HDM}} & = & 
- \sum_{i,j=1}^3 {\Big \lbrace}
{\tilde D}_{ij}^{(1)}
( {\bar {\tilde q}}^{(i)}_L {\Phi}^{(1)} )
{\tilde d}^{(j)}_{R} +
{\tilde D}_{ij}^{(2)}
( {\bar {\tilde q}}^{(i)}_L {\Phi}^{(2)} )
{\tilde d}^{(j)}_{R} +
    \nonumber\\
& & + {\tilde U}_{ij}^{(1)}
( {\bar {\tilde q}}^{(i)}_L \tilde {\Phi}^{(1)} )
{\tilde u}^{(j)}_{R} +
{\tilde U}_{ij}^{(2)}
( {\bar {\tilde q}}^{(i)}_L \tilde {\Phi}^{(2)} )
{\tilde u}^{(j)}_{R} + {\rm{h.c.}}
{\Big \rbrace} \ .
\label{2HD30}
\end{eqnarray}
Parameters and quark fields are in an arbitrary 
$\rm{SU(2)_L}$-basis, as indicated by tildes above them.
Superscript $(E)$ at the Lagrangian density means that
the theory has a finite effective energy cutoff $E$
(energy of probes).
This superscript is omitted at the fields and at the
parameters for simpler notation.\footnote{
For renormalized quantities, $E\!\sim\!10^2$ GeV.} 
The following notations are used:
\begin{equation}
{\Phi}^{(k)}  \equiv  
{ {\phi}^{(k)+} \choose {\phi}^{(k)0} } 
\equiv \frac{1}{\sqrt{2}} 
{ {\phi}_1^{(k)} + {\rm{i}} {\phi}_2^{(k)} \choose
     {\phi}_3^{(k)} + {\rm{i}} {\phi}_4^{(k)} }  \ , \quad
{\tilde {\Phi}}^{(k)}  \equiv 
{\rm{i}} {\tau}_2 {\Phi}^{(k)\ast} \ ,
\label{2HDnot1}
\end{equation}
\begin{equation}
{\tilde q^{(i)}} = 
{ {\tilde u^{(i)}} \choose {\tilde d^{(i)}} } \ :
\qquad
{\tilde q^{(1)}} = { {\tilde u} \choose {\tilde d} } \ , \
{\tilde q^{(2)}} = { {\tilde c} \choose {\tilde s} } \ , \
{\tilde q^{(3)}} = { {\tilde t} \choose {\tilde b} } \ ,
\label{2HDnot2}
\end{equation}
\begin{equation}
\langle {\Phi}^{(1)} \rangle_0 = 
\frac{{\rm{e}}^{{\rm{i}}\eta_1}}{\sqrt{2}} 
{0 \choose v_1} \ ,
\qquad
\langle {\Phi}^{(2)} \rangle_0 = \
\frac{{\rm{e}}^{{\rm{i}}\eta_2}}{\sqrt{2}} 
{0 \choose v_2} \ ,
\qquad v_1^2+v_2^2 = v^2 \ .
\label{2HDnot3}
\end{equation}
Here, $v$ [$\equiv v(E)$] is the usual vacuum expectation 
value (VEV) needed for the electroweak symmetry breaking:
$v(E_{\rm{ew}})\!\approx\!246$ GeV. 
Phase difference $\eta\!\equiv\!\eta_2\!-\!\eta_1$ 
between the two VEV's in (\ref{2HDnot3}) may be nonzero 
and represents CP violation originating from the purely 
scalar sector.

Popular 2HDM(I) and 2HDM(II) models proposed by
Glashow and Weinberg \cite{GlashowWeinberg}
are special cases (subsets)
of this framework: $U^{(1)}\!=\!D^{(1)}\!=\!0$ in 2HDM(I), 
and $U^{(1)}\!=\!D^{(2)}\!=\!0$ in 2HDM(II).
In these models, mass matrices $M^{(U)}$
and $M^{(D)}$ in the mass basis 
are proportional to the relevant nonzero Yukawa
matrices $U$, $D$, respectively. Therefore, there are no
FCNYCP's in the physical (quark mass) basis,
since off-diagonal elements are zero there by definition.
Moreover, this remains true even when radiative
corrections are included, i.e., when the model is
evolved from a high ``bare'' energy $E\!=\!\Lambda$
to low energies $E\!\sim\!m_q$. This is so because
the mentioned structure of the Yukawa sector
is ensured by specific discrete symmetries. 

There are no such symmetries in the
Yukawa sector of G2HDM, so a priori it is unclear
whether suppression of FCNYCP's would persist
when we increase energy of probes beyond electroweak
scales. We carried out such an analysis, by using
one--loop renormalization group equations (RGE's)
for Yukawa coupling matrices appearing in (\ref{2HD30}).
Those RGE's were derived by us earlier \cite{CHK}.

In order to interpret more easily the RGE results 
at high energies of probes, as well as the initial 
conditions at low (electroweak) scales, it is 
convenient to introduce the following derived quantities:
\begin{equation}
{\Phi}^{\prime(1)} = \cos \beta {\Phi}^{(1)} +
{\rm{e}}^{-{\rm{i}} {\eta}} \sin \beta {\Phi}^{(2)},
\quad
{\Phi}^{\prime(2)} = - \sin \beta {\Phi}^{(1)} +
{\rm{e}}^{-{\rm{i}} {\eta}} \cos \beta {\Phi}^{(2)},
\label{redefPhi}
\end{equation}
\begin{equation}
\mbox{where: } \ 
\tan \beta = \frac{v_2}{v_1} \ \Rightarrow \
\cos \beta = \frac{v_1}{v} \ , 
\quad \sin \beta = \frac{v_2}{v} \ ;
\quad \eta=\eta_2-\eta_1 \ .
\label{ratioVEV}
\end{equation}
VEV's of the redefined scalar isodoublets are:
${\rm{e}}^{-{\rm{i}} {\eta}_1} 
\langle {\Phi}^{\prime(1)} \rangle_0^T\!=\!(0, v/\sqrt{2})$,
$\langle {\Phi}^{\prime(2)} \rangle_0^T\!=\!(0,0)$.
Hence, isodoublet ${\Phi}^{\prime(1)}$ is responsible
for the quark masses. Below we will see 
that ${\Phi}^{\prime(2)}$ leads to FCNYCP's.
We now rewrite original Yukawa Lagrangian density 
(\ref{2HD30}) of G2HDM in terms of these fields
\begin{eqnarray}
{\cal L}^{(E)}_{\rm{G2HDM}} & = & 
- \sum_{i,j=1}^3 {\Big \lbrace}
{\tilde G}^{(D)}_{ij}
( {\bar {\tilde q}}^{(i)}_L {\Phi}^{\prime (1)} )
{\tilde d}^{(j)}_{R} +
{\tilde G}^{(U)}_{ij}
( {\bar {\tilde q}}^{(i)}_L {\tilde \Phi}^{\prime(1)} )
{\tilde u}^{(j)}_{R} + {\rm{h.c. }} {\Big \rbrace}
\nonumber\\
&&
- \sum_{i,j=1}^3 {\Big \lbrace}
{\tilde D}_{ij}
( {\bar {\tilde q}}^{(i)}_L {\Phi}^{\prime (2)} )
{\tilde d}^{(j)}_{R} +
{\tilde U}_{ij}
( {\bar {\tilde q}}^{(i)}_L {\tilde \Phi}^{\prime(2)} )
{\tilde u}^{(j)}_{R} + {\rm{h.c. }} {\Big \rbrace} \ .
\label{Lnew}
\end{eqnarray}
${\tilde G}^{(U)}$ and ${\tilde G}^{(D)}$ are rescaled 
mass matrices, and ${\tilde U}$ and ${\tilde D}$
the corresponding Yukawa matrices ``orthonormal'' to
${\tilde G}^{(U)}$ and ${\tilde G}^{(D)}$
\begin{eqnarray}
{\tilde G}^{(X)} &=& \cos \beta {\tilde X}^{(1)} + 
{\rm{e}}^{\mp{\rm{i}} {\eta}} \sin \beta {\tilde X}^{(2)} 
= {\sqrt{2}} {\tilde M}^{(X)}/v  \quad (X=U,D) \ ,
\label{Gs}
\\
{\tilde X} &=& - \sin \beta {\tilde X}^{(1)} +
{\rm{e}}^{\mp{\rm{i}}{\eta}} \cos \beta {\tilde X}^{(2)} 
\qquad (X=U,D) \ .
\label{UDs}
\end{eqnarray}
Minus sign in exponents is for $X\!=\!U$, and plus for 
$X\!=\!D$. Transition to the quark mass basis (at a given
energy $E$) is implemented by biunitary transformations
involving unitary matrices $V^U_{L,R}$ and $V^D_{L,R}$
\begin{eqnarray}
U = V_L^{U} {\tilde U} V_R^{U\dagger} ; \quad
G^{(U)} &=& V_L^{U} {\tilde G}^{(U)} V_R^{U\dagger} , 
\quad G_{ij}^{(U)}= {\delta}_{ij} m_i^{(u)} {\sqrt{2}}/v \ ;
\label{GUUmass}
\\
D = V_L^{D} {\tilde D} V_R^{D\dagger} ; \quad
G^{(D)} &=& V_L^{D} {\tilde G}^{(D)} V_R^{D\dagger} ,
\quad G_{ij}^{(D)}= {\delta}_{ij} m_i^{(d)} {\sqrt{2}}/v \ ;
\label{GDDmass}
\\
u_L =V_L^{U} {\tilde u}_L , \quad 
u_R &=& V_R^U {\tilde u}_R , \quad
d_L = V_L^{D} {\tilde d}_L , \quad 
d_R = V_R^D {\tilde d}_R \ .
\label{massbasis}
\end{eqnarray}
Parameters and fields without tildes are in the mass basis. 
Superscripts ($E$) are omitted for simpler notation.
CKM mixing matrix is $V\!\equiv\!V_L^{U}V_L^{D\dagger}$.
Neutral part of Lagrangian density (\ref{Lnew}) in the
quark mass basis is then
\begin{eqnarray}
\lefteqn{\! \! \! \! \! \! \! \! \! \!
{\cal L}^{(E) \rm{neutr.}}_{\rm{G2HDM}} =
- \frac{1}{\sqrt{2}} {\Big \lbrace}
G^{(D)}_{ii} {\bar d}^{(i)}_L d^{(i)}_{R} 
[{\phi}^{\prime(1)}_3\!+\!{\rm{i}} {\phi}^{\prime(1)}_4] +
G^{(U)}_{ii} {\bar u}^{(i)}_L u^{(i)}_{R}
[{\phi}^{\prime(1)}_3\!-\!{\rm{i}} {\phi}^{\prime(1)}_4]}
\nonumber\\
&&
+ D_{ij} {\bar d}^{(i)}_L d^{(j)}_{R} 
[{\phi}^{\prime(2)}_3\!+\!{\rm{i}} {\phi}^{\prime(2)}_4] +  
U_{ij} {\bar u}^{(i)}_L u^{(j)}_{R}
[{\phi}^{\prime(2)}_3\!-\!{\rm{i}} {\phi}^{\prime(2)}_4]
 + {\rm{h.c.}} {\Big \rbrace} \ ,
\label{Lmassn}
\end{eqnarray}
where summation is performed over repeated flavor indices.
We see from (\ref{Lmassn}) that the $U$ and $D$ matrices,
defined by (\ref{UDs})--(\ref{GDDmass})
through the original matrices ${\tilde U}^{(j)}$ and
${\tilde D}^{(j)}$ of  G2HDM (\ref{2HD30}), allow the model 
to possess in general FCNYCP's. This is so because
in the quark mass basis only the
(rescaled) quark mass matrices $G^{(U)}$ and $G^{(D)}$ 
of (\ref{GUUmass})-(\ref{GDDmass}) [cf.~also (\ref{Gs})]
are diagonal, while additional matrices $U$ and $D$ 
are in general not.

As mentioned earlier, CSY \cite{CSY} argued that
$U$ and $D$ matrices in the quark 
mass basis and at low energies $E$ have the form
\begin{equation}
U_{ij}(E) = {\xi}^{(u)}_{ij} \frac{\sqrt{2}}{v}
\sqrt{m_i^{(u)} m_j^{(u)}} \ , \qquad
D_{ij}(E) = {\xi}^{(d)}_{ij} \frac{\sqrt{2}}{v}
\sqrt{m_i^{(d)} m_j^{(d)}} \ , 
\label{FCNCcon1}
\end{equation}
\begin{equation}
\mbox{with: } \qquad 
{\xi}^{(u)}_{ij}, {\xi}^{(d)}_{ij} \sim 1 
\quad {\rm{for }} \ 
E \sim E_{\rm{ew}} (\sim M_Z) \ .
\label{FCNCcon2}
\end{equation}
This form, at least for the diagonal elements,
is suggested by the existing quark mass
hierarchies and the requirement that there be
no fine--tuning in which large Yukawa terms
$U^{(i)}_{jk}$ (and: $D^{(i)}_{jk}$)
would add together via (\ref{UDs})
to result in much smaller terms $U_{jk}$ ($D_{jk}$) --
for this, inspect Eqs.~(\ref{Gs})--(\ref{UDs}), but this
time in the quark mass basis (no tildes).
Moreover, this ansatz turns out to be phenomenologically
viable, i.e., the known suppression of flavor--changing
neutral processes at low energies (not involving 
on--shell top quarks) can be reproduced.
Similar (but not identical) ans\"atze have been proposed by
the authors of \cite{AHR} (AHR),
motivated by their requirement that the Yukawa interactions
have approximate $U(1)$ flavor symmetries.

Now we are prepared to present some typical examples
of the RGE evolution of Yukawa parameters in G2HDM's.
For simplicity of analysis, we neglect contributions of
the first quark generation. Further,
we assume that all original four Yukawa matrices 
${\tilde U}^{(j)}$, ${\tilde D}^{(j)}$ are real and the
VEV phase difference $\eta$ is zero (no CP violation).

For the boundary conditions to the RGE's, at
the evolution energy $E=M_Z$,
we first took the CSY ansatz (\ref{FCNCcon1})-(\ref{FCNCcon2}),
with $\xi^{(u)}_{ij}\!=\!1\!=\!\xi^{(d)}_{ij}$ or
$\xi^{(u)}_{ij}\!=\!2\!=\!\xi^{(d)}_{ij}$,
for all $i,\!j\!=\!1,\!2$. We emphasize 
that $i\!=\!1$ refers now to the second quark family 
({\em c,s\/}), and $i\!=\!2$ 
to the third family ({\em t,b\/}). 
For the ($2\!\times\!2$) 
orthonormal CKM mixing matrix $V$ we take
$V_{12}(M_Z)\!=\!0.045\!=\!-V_{21}(M_Z)$. Values of other
parameters at $E\!=\!M_Z$ were chosen to be:\\
$\tan \beta\!=\!1.0$; 
$v\!\equiv\!\sqrt{v_1^2+v_2^2}\!=\!246.22$ GeV;
$\alpha_3\!=\!0.118$, $\alpha_2\!=\!0.332$, 
$\alpha_1\!=\!0.101$; 
$m_c\!=\!0.77$ GeV, $m_s\!=\!0.11$ GeV, 
$m_b\!=\!3.2$ GeV, and $m_t\!=\!171.5$ GeV.
These quark mass values correspond to: 
$m_c(m_c)\!\approx\!1.3$ GeV,
$m_s(1 {\rm{GeV}})$ $\approx 0.2$ GeV, 
$m_b(m_b)\!\approx\!4.3$ GeV,
and $m_t^{\rm{phys.}}\!\approx\!174$ GeV
[$m_t(m_t)\!\approx\!166$ GeV]. For ${\alpha}_3(E)$ we
used two-loop evolution formulas, with threshold effect
at $E\!\approx\!m_t^{\rm{phys.}}$ taken into
account; for ${\alpha}_j(E)$ ($j\!=\!1,\!2$) 
we used one-loop evolution formulas. 

This simplified framework resulted in $18$ coupled
RGE's [for $18$ real parameters: $v^2$, $\tan \beta$,
${\tilde U}_{ij}$, ${\tilde D}_{ij}$, 
${\tilde G}^{(U)}_{ij}$, ${\tilde G}^{(D)}_{ij}$], 
with the mentioned boundary conditions at $E\!=\!M_Z$.
The RGE system was solved numerically, using
Runge-Kutta subroutines with adaptive stepsize control
(given in \cite{WHPressetal}). 

The results for the ratios of FCNYCP's
$X_{ij}(E)/X_{ij}(M_Z)$
($X\!=\!U,\!D$; $i\!\not=\!j$) 
are given for the case of
$\xi^{(u)}_{ij}\!=\!1\!=\!\xi^{(d)}_{ij}$ in Fig.~1.
\begin{figure}[htb]
\mbox{}
\vskip10.5cm\relax\noindent\hskip1.1cm\relax
\includegraphics{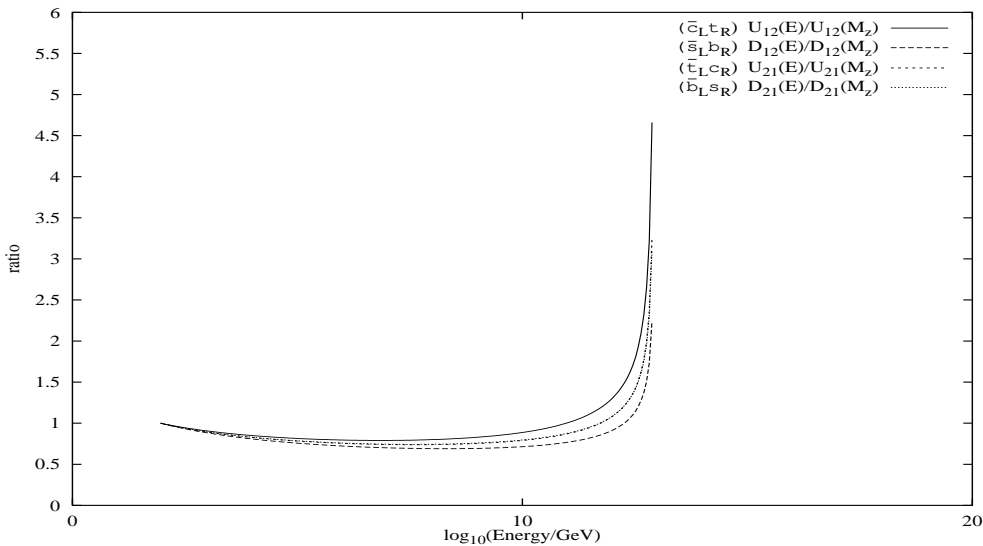} \vskip-3.7cm
\caption{\footnotesize Evolution of FCNYCP ratios 
$U_{ij}(E)/U_{ij}(M_Z)$, 
$D_{ij}(E)/D_{ij}(M_Z)$ ($i\!\not=\!j$),
when ${\xi}^{(u)}_{ij}\!=\!{\xi}^{(d)}_{ij}\!=\!1$
(for all $i,\!j\!=\!1,\!2$).}
\end{figure}
The Figure shows that all the FCNYCP's
are remarkably stable as the evolution energy
increases. Even the up-type FCNYCP's, appearing in 
couplings involving the heavy $t$ quark, 
remain rather stable. 
Only very close to the $t$-quark-dominated
Landau pole ($\sim$$10^{13}$ GeV)
these parameters start increasing substantially.
For example, in the down-type FCN sector ($b$-$c$)
the corresponding ratio $D_{21}(E)/D_{21}(M_Z)$
acquires its double initial value (i.e., value $2$)
at $E\!\approx\!0.7 E_{\rm{pole}}$, which
is very near the pole.
About the same is true also for
$U_{21}(E)/U_{21}(M_Z)$. 
For the ratio $D_{12}(E)/D_{12}(M_Z)$
the corresponding energy is even closer to 
$E_{\rm{pole}}$, while for for the $t$--quark--dominated
$U_{12}(E)/U_{12}(M_Z)$ it is somewhat lower.
We draw the conclusion that
the case $\xi^{(u)}_{ij}\!=\!1\!=\!\xi^{(d)}_{ij}$
shows no ``unnaturality'' 
concerning the behavior of FCNYCP's at high energies.
We may also compare this stability with that of other ratios. 
For example, in Ref.~\cite{CHK} we showed
also behavior of those ratios of neutral Yukawa
parameters which don't involve
flavor--changing couplings: 
$U_{jj}(E)/U_{jj}(M_Z)$ and $D_{jj}(E)/D_{jj}(M_Z)$;
$G^{(U)}_{jj}(E)/G^{(U)}_{jj}(M_Z)$ and 
$G^{(D)}_{jj}(E)/G^{(D)}_{jj}(M_Z)$.
We saw that several (but not all) of these coupling 
parameters are also reasonably stable. The results for 
FCNYCP's of Fig.~1 are interesting and perhaps surprising.
We should bear in mind that the off--diagonal Yukawa 
parameters appearing in Fig.~1 are at low energies 
by several factors smaller than the third generation 
diagonal parameters, due to CSY ansatz.\footnote{
Note that $D_{12}\!=\!D_{21}$ at $E\!=\!M_Z$
is two orders of magnitude smaller than
$U_{22}$.}
Therefore, the fear that the latter, substantially larger, 
parameters would ``pull up'' the suppressed 
off--diagonal ones (by a large factor,
or even by orders of magnitude) as the 
energy of probes increases, is
intuitively justified. Fig.~1 says that this doesn't happen.

One may object that behavior shown in Fig.~1
is due to a special choice 
$\xi^{(u)}_{ij}\!=\!1\!=\!\xi^{(d)}_{ij}$. In fact, in
Ref.~\cite{CHK}, where the relevant RGE's were derived,
we performed numerical analysis only
for the case ${\xi}^{(u)}_{ij}\!=1\!=\!{\xi}^{(d)}_{ij}$.
We have recently performed calculations for 
variations of the CSY ansatz.
Results are independent of the VEV ratio
$\tan \beta$ (cf.~also discussion in \cite{CHK}).
\begin{figure}[htb]
\mbox{}
\vskip10.1cm\relax\noindent\hskip1.5cm\relax
\includegraphics{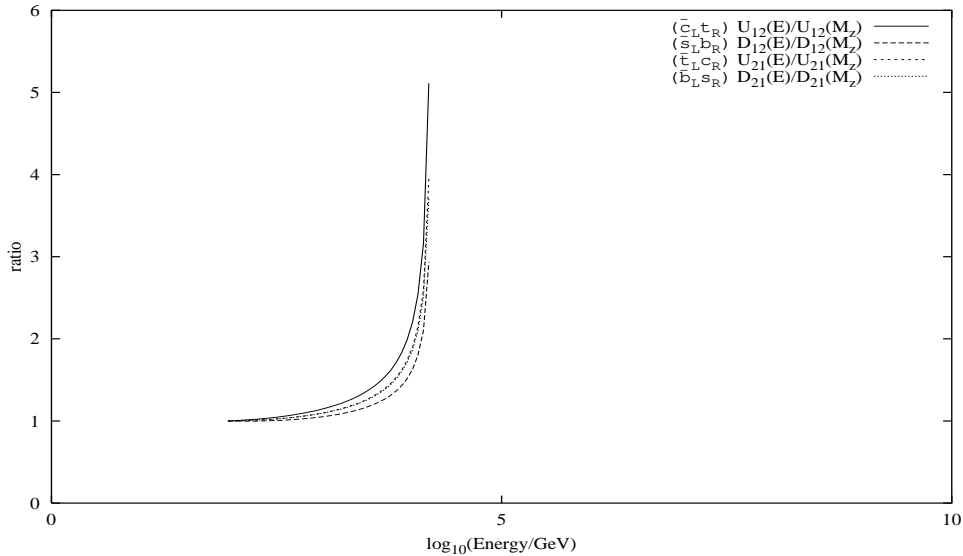} \vskip-3.4cm
\caption{\footnotesize Same as in Fig.~1, but for the
choice ${\xi}_{ij}^{(u)}\!=\!{\xi}_{ij}^{(d)}\!=2$
(for all $i,\!j\!=\!1,\!2$).}
\end{figure}
In Fig.~2 we show results when
$\xi^{(u)}_{ij}\!=\!2\!=\!\xi^{(d)}_{ij}$ at 
$E\!=\!M_Z$ (for all $i,\!j\!=\!1,\!2$).
The Landau pole is now of course
lower ($\sim$$10$ TeV), but the behavior 
with respect to the stability of FCNYCP's remains 
qualitatively the same. When we vary
some of the $\xi_{ij}$ parameters, the results remain
qualitatively the same, while the Landau pole is
influenced largely by the ($t$--quark--dominated)
up-type CSY parameter ${\xi}_{22}^{(u)}$ [$U_{22}(M_Z)$,
cf.~Eqs.~(\ref{FCNCcon1})--(\ref{FCNCcon2})]. 
We investigated also cases which go beyond CSY ansatz.
For example, we chose to suppress the up-type 
off--diagonal parameter even more drastically, by taking
${\xi}_{12}^{(u)}\!={\xi}_{21}^{(u)}\!=\!0.05163$ 
and all other ${\xi}_{ij}$ parameters equal to $1$.
For such a choice, we have 
$D_{12}\!=\!D_{21}\!=\!U_{12}\!=\!U_{21}$ at $E\!=\!M_Z$.
The results are given in Fig.~3, and are very close to those
of Fig.~1.
\begin{figure}[htb]
\mbox{}
\vskip10.1cm\relax\noindent\hskip1.5cm\relax
\includegraphics{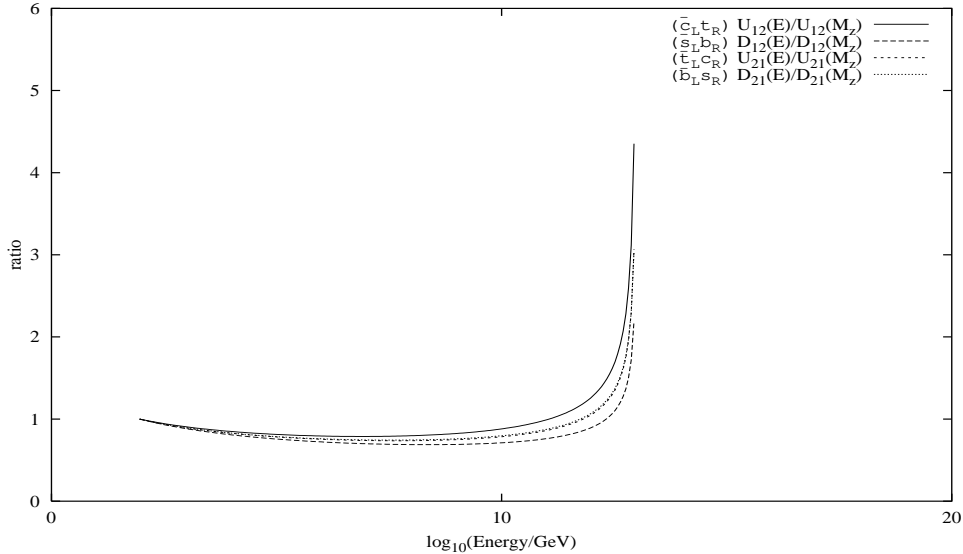} \vskip-3.4cm
\caption{\footnotesize Same as in Fig.~1, but
for the choice ${\xi}_{12}^{(u)}\!=\!{\xi}_{21}^{(u)}\!=\!0.05163$
(other ${\xi}_{ij}$'s are $1$).}
\end{figure}

We conclude that in G2HDM's, with
effects of the first light family neglected,
the flavor--changing neutral Yukawa coupling parameters
(FCNYCP's) remain remarkably stable as the energy of
probes increases up to the vicinity of the
(top--quark--dominated) Landau pole. The conclusion
is insensitive to any specific
variation of the CSY ansatz (\ref{FCNCcon1})--(\ref{FCNCcon2})
for Yukawa parameters at low energies,
and apparently survives even when going beyond this ansatz
by suppressing low energy FCNYCP's even more.
Thus, in the presented framework, the fear that the
suppressed FCNYCP's at low energies are pulled up
drastically by the much larger diagonal Yukawa coupling
parameters as the energy of probes increases doesn't
materialize. The described framework appears to behave
naturally in this respect, FCNYCP's remain remarkably
suppressed even at higher energies although there is
no explicit exact symmetry which would ensure complete
suppression of these parameters. In this sense, we
have an indication that the G2HDM's
are not unnatural, in contrast with the widely held
beliefs.

One may still argue that our conclusions might change when
contributions of the first quark generation are included. 
We intend to perform this
extension of numerical analysis in the near future.

\vspace{0.5cm}

\noindent {\bf Abbreviations frequently used in the article:}

\noindent CSY -- Cheng, Sher and Yuan;
FCNC -- flavor-changing neutral currents;
FCNYCP -- flavor-changing neutral Yukawa coupling parameter;
G2HDM -- general two-Higgs-doublet (Standard) Model;
RGE -- renormalization group equation;
VEV -- vacuum expectation value.

\newpage

\centerline{{\bf Acknowledgments}}

\vspace{0.3cm}

\noindent The work of CSK and SSH was supported 
in part by the CTP of SNU, 
in part by Yonsei University Faculty Research Fund of 1997, 
in part by the BSRI Program, Ministry of Education, 
Project No. BSRI-97-2425, 
in part by Non-Directed-Research-Fund of 1997, KRF, 
and in part by the KOSEF-DFG large collaboration
project, Project No. 96-0702-01-01-2.
GC thanks Prof.~Schildknecht for
offering him financial support of Bielefeld University
during the course of this work.

\end{document}